\documentclass[preprint,showpacs,preprintnumbers,amsmath,amssymb,natbib]{revtex4}
\usepackage{graphicx}% Include figure files
\usepackage{epsfig}		
\usepackage{dcolumn}% Align table columns on decimal point
\usepackage{bm}% bold math
\usepackage{subfigure}

\def\0{\mbox{\tiny $0$}}
\def\1{\mbox{\tiny $1$}}
\def\2{\mbox{\tiny $2$}}
\def\3{\mbox{\tiny $3$}}
\def\4{\mbox{\tiny $4$}}
\def\5{\mbox{\tiny $5$}}
\def\6{\mbox{\tiny $6$}}
\def\7{\mbox{\tiny $7$}}
\def\8{\mbox{\tiny $8$}}
\def\9{\mbox{\tiny $9$}}

\def\f14{\mbox{\tiny $\frac{1}{4}$}}

\begin{document}

\title{Quantum engines and the range of the second law of thermodynamics in the noncommutative phase-space}

\author{Jonas F. G. Santos and Alex E. Bernardini\footnote{E-mail: jonas@df.ufscar.br, alexeb@ufscar.br}} 
\affiliation{Departamento de F\'isica, Universidade Federal de S\~ao Carlos, PO Box 676, 13565-905, S\~ao Carlos, SP, Brasil.}

\date{\today}
\pacs{05.30.Ch, 05.70.-a }

\begin{abstract}

Two testable schemes for quantum heat engines are investigated under the quantization framework of noncommutative (NC) quantum mechanics (QM). By identifying the phenomenological connection between the phase-space NC driving parameters and an effective external magnetic field, the NC effects on the efficiency coefficient, $\mathcal{N}$, of quantum engines can be quantified for two different cycles: an isomagnetic one and an isoenergetic one.
In addition, paying a special attention to the quantum Carnot cycle, one notices that the inclusion of NC effects does not affect the maximal (Carnot) efficiency, $\mathcal{N}^C$, ratifying the robustness of the second law of thermodynamics. 
\end{abstract}

\maketitle

\section{Introduction}

The manipulation of energy stored by nano-devices driven by quantum heat engine schemes has raised some of the ultimate nano-technological achievements to a novel and challenging baseline \cite{Novo01,Novo02,Novo03,Novo04}.
The experimental engineering of thermodynamical nano-cycles has been adapted for emulating quantum heat engines driven by trapped ions \cite{3000}, for detecting the effect of local quantum correlations on the formulation of the thermodynamic axioms \cite{Novo01,Novo00,Novo03},
for building opto-mechanical systems \cite{5000}, and also for reproducing the principles of photosynthesis through photo-Carnot engines \cite{6000,7000}.
Quantum mechanical versions of the engine cycles \cite{1300,1400,Novo02}, photosynthesis emulated by photocells \cite{1000,1100,1200} and controllable squeezed thermal states \cite{1400} are all encompassed by the modern range of implementations of quantum heat engines through nano-devices.

On the theoretical front, an imperious question arises when the underlying properties imposed by quantum mechanics affect the involved thermodynamic processes \cite{Novo01,Novo00,Novo0B}.
In particular, the heat engine maximal efficiency derived from the Carnot statement of the second law of thermodynamics has been discussed in the context of the inclusion of (additional) quantum correlational approaches \cite{Novo02}.
Nevertheless, due to finite-size effects and overall limitations of tiny systems with only a few degrees of freedom, as well as to some involved inaccessible forms of heat or chemical energy, the dramatical influence of quantum effects on thermodynamic properties still misses a consensual point.

From a complemental perspective, our aim is to generalize the investigation of quantum thermodynamic cycles and quantum heat engines to the scope of noncommutative (NC) extensions of quantum mechanics (QM) \cite{Bernardini01,Catarina,Catarina001}.
Looking for an encompassing correspondence between quantum and classical systems \cite{Catarina002}, several issues on NC QM related to Gaussian quantum correlations \cite{Bernardini13B,Bernardini13B2}, quantum coherence and information collapse \cite{Bernardini01,2015,2016}, and violations of the uncertainty relation in QM \cite{Catarina001,Bernardini13C,Bernardini13D,Bernardini13E} have been recently examined.
From the perspective of a deformed Heisenberg-Weyl algebra of QM \cite{Rosenbaum,Gamboa}, which is realized by the commutation relations given by
\begin{equation}
\left[\hat{q}_i,  \hat{q}_j \right] = i \theta_{ij} , \qquad \left[ \hat{q}_i,  \hat{p}_j \right] = i \hbar \delta_{ij} ,
\qquad \left[ \hat{p}_i,  \hat{p}_j \right] = i \eta_{ij} ,  \qquad i,j= 1, ... ,D
\label{Ape001}
\end{equation}
in $D$ dimensions, where $\eta_{ij}$ and $\theta_{ij}$ are invertible antisymmetric real constants, the NC QM has supported modified versions of Landau level and $2D$ harmonic oscillator problems in the phase-space \cite{Nekrasov01,Rosenbaum}, corrections to the quantum Hall effect \cite{Prange}, and also the quantization of the gravitational well for cold neutrons \cite{Bertolami01,Bertolami02,Banerjee}. 
On a still broader context, noncommutativity is also believed to answer some fundamental questions in quantum gravity and string theory \cite{Connes,Douglas,Seiberg}.

The proposal of this work is to extend the investigation of thermodynamic cycles to quantum heat engines embedded into the framework of NC QM. 
Since the quantum thermodynamic scenarios for simplified quantum heat engines involves a reduced number of degrees of freedom, one may assume that the natural emergence of NC parameters into the calculations, and the corresponding discretization of NC QM effects can be easily identified and possibly connected to the phenomenological predictions.

Considering the ordinary QM architecture of single-particle quantum heat engines, which is confined by a (quasi-statically) controllable axially symmetric magnetic field potential combined with a cylindrical potential well (c. f. Refs. \cite{Novo02}), one can quantify the corresponding modifications introduced by the NC deformation of QM.
Once it has been performed, one can follow the thermodynamic heat/work calculation protocol as to obtain the efficiency, $\mathcal{N}$, of the two most relevant schemes for quantum engines.

Given the above mentioned proposal, the paper is organized as follows.
In Section II, the theoretical preliminaries for describing quantum engine cycles are introduced.
The Section III brings-up the main results of this work.
The so-called isomagnetic and isoenergetic thermodynamic schemes are both investigated and generalized to the phase-space NC framework.
The former one is composed by two isomagnetic and two isoentropic trajectories, and the latter one by two isoentropic and two isoenergetic trajectories, both reversible. 
Such quantum engines are driven by the intensity of the external magnetic field, which is the relevant manipulable degree of freedom for the quantum schemes. 
Finally, the impact of NC effects on the quantum Carnot cycle is identified and quantified for the NC scenario of QM supported by the Heisenberg-Weyl algebra from Eq.~(\ref{Ape001}).
Our conclusions are drawn in Section IV.

\section{Theoretical Preliminaries}

Let one considers a quantum system composed by a single-particle confined by a cylindrical potential well of characteristic frequency $\omega$, and on the influence of an external magnetic field along the $z$ axis,
\begin{equation}
\textbf{B} = B \hat{z},
\end{equation}
defined through an adopted symmetric gauge
\begin{equation}
\textbf{A} = \frac{B}{2}(-q_2, q_1, 0).
\end{equation}

The Hamiltonian of the system is built as to be read as
\begin{equation}
\hat{H}(q_i,\, p_i) = \frac{p_1^2 + p_2^2}{2m} + \frac{m \Omega^2}{2}(q_1^2 + q_2^2) + \frac{\omega_{_B}}{2}(q_2\, p_1 - q_1\, p_2),
\label{Hamil}
\end{equation}
where $m$ is the mass of the particle, $\omega_{_B}  = q\, B/ m$ is the associated frequency of the magnetic field, and an effective frequency can be identified by
\begin{equation}
\Omega^2 = \frac{\omega_{_B}^2}{4} + \omega^2.
\end{equation}

It must be reinforced that the choice of a cylindrical potential to confine the particle works as a good approximation of a confinement scheme in semiconductor quantum dots \cite{Jacak}. In fact, quantum dots with a spin orbit interaction has already been studied in the context of a NC plane in \cite{Pal}, where it has been shown that the NC effect can also be modulated as an external magnetic field.

Once the Heisenberg-Weyl algebra is deformed into its NC version described by the Eq.~(\ref{Ape001}), the NC operators, $(q_{1,2}, p_{1,2})$, which define $\hat{H}(q_i,\, p_i)$, can be mapped onto the ordinary Hilbert space through the following Seiberg-Witten correspondence (see the Appendix) \cite{Seiberg},
\begin{eqnarray}
\hat{q}_1 &=& \nu\, \hat{Q}_1 - \frac{\theta}{2\nu \hbar}\hat{\Pi}_2, \quad \hat{p}_1 = \mu \hat{\Pi}_1 + \frac{\eta}{2\mu \hbar}\hat{Q}_2,\nonumber\\
\hat{q}_2 &=& \nu\, \hat{Q}_2 + \frac{\theta}{2\nu \hbar}\hat{\Pi}_1, \quad \hat{p}_2 = \mu \hat{\Pi}_2 - \frac{\eta}{2\mu \hbar}\hat{Q}_1,
\label{sssw}
\end{eqnarray} 
where $\mu$ and $\nu$ are arbitrary mapping constants, and $\theta$ and $\eta$ are the NC parameters identified with $\theta_{ij}$ and $\eta_{ij}$ from Eq.~(\ref{Ape001}) in a $2$-dim scenario.
After some mathematical manipulations, the map from (\ref{sssw}) leads to an ordinary (standard) QM version of the Hamiltonian,
\begin{equation}
\hat{H}(Q_i,\, \Pi_i) = \tilde{\alpha}^2(Q_1^2 + Q_2^2) + \tilde{\beta}^2(\Pi_1^2 + \Pi_2^2) + \left(\frac{\omega_{_B}}{2} + \gamma\right)(\Pi_1 Q_2 - \Pi_2 Q_1),
\label{Hamilt02}
\end{equation}
where
\begin{eqnarray}
\tilde{\alpha}^2 &=& \frac{\nu^2 m \Omega^2}{2} + \frac{\eta^2}{8m\mu^2 \hbar^2} + \frac{\nu}{\mu}\frac{\omega_{_B} \eta}{4\hbar},\nonumber\\
\tilde{\beta}^2 &=& \frac{\mu^2}{2m} + \frac{m \Omega^2 \theta^2}{8\nu^2 \hbar^2} + \frac{\mu}{\nu}\frac{\omega_{_B} \theta}{4	\hbar}.\nonumber\\
\gamma &=& \frac{\theta}{2\hbar}m\Omega^2 + \frac{\eta}{2m\hbar}.
\label{values}
\end{eqnarray}

The Hamiltonian from (\ref{Hamilt02}) is supported by the \textit{eigenvalue} equation for QM and admits a wave function given in terms of associated Laguerre polynomials with the following \textit{eigenenergies} \cite{Bernardini01}
\begin{equation}
E_{\kappa,\, \ell} = 2\hbar\, \tilde{\alpha}\tilde{\beta}(2\, \kappa + |\ell| + 1) - \hbar\left(\frac{\omega_{_B}}{2} + \gamma\right)\ell.
\label{Eingenva}
\end{equation}

\begin{figure}
%\hspace{1.0 cm}
\centering
\includegraphics[scale=0.50]{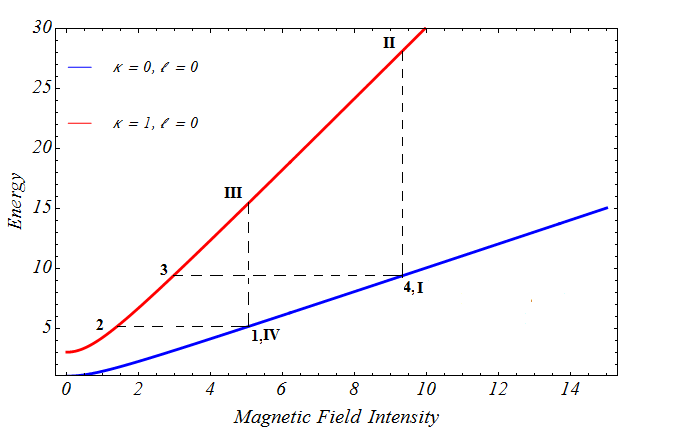}
\caption{(Color online) Isomagnetic (Roman indices) and isoenergetic (Latin indices) cycles for the effective two level system engendred by the ground state, $(\kappa = 0, \ell = 0)$, and by the first excited state, $(\kappa = 1, \ell = 0)$. The isomagnetic cycle involves two isomagnetic and two isoentropic trajectories, whereas the isoenergetic cycle involves two isoenergetic and two isoentropic ones.}
\label{Scheme}
\end{figure}

Once the idealized quantum heat engine works through two different quantum states as shown in Fig. \ref{Scheme}, it is convenient to re-write the \textit{eingenenergies} as functions of the NC parameters, as to quantify their influence on the quantum cycles. By replacing the relations from (\ref{values}) into (\ref{Eingenva}) one gets
\begin{equation}
E_{\kappa,\, \ell} = \sigma\, \hbar\, \omega\,\sqrt{1 + \mathcal{F}_\Phi^2}\,(2\kappa + |\ell| + 1) - \hbar \left(\frac{\omega_{_B}}{2} + \gamma\right)\ell,
\label{Eingelvalues}
\end{equation}
where $\mathcal{F}_\Phi^2$ is identified by
\begin{equation}
\mathcal{F}_\Phi^2 = \frac{1}{\sigma^2 \omega^2}\left(\frac{\omega_{_B}^2}{4} + \gamma\omega_{_B} + \gamma^2\right),
\end{equation}
and $\sigma$ is a function of the NC parameters, $\theta$ and $\eta$,
\begin{equation}
\sigma = \sqrt{1 - \frac{\theta \eta}{\hbar^2}}.
\end{equation}

By observing an eventual vanishing behavior of $\gamma$ and $\sigma$, which reduces the NC deformed algebra to the standard Heisenberg-Weyl algebra, the factor $\mathcal{F}_\Phi^2$ goes to $N_{\Phi} = \omega_{_B}/2\omega$, which has already been identified by the Refs. \cite{Novo02} as the flux quanta. 
Therefore, the novel flux quanta here obtained exhibits both magnetic and NC properties.

To guarantee the quasi-static variation of the magnetic field intensity as the unique driving force able to induce quantum transitions along the schemes from Fig.~\ref{Scheme}, one has to impose the selection rule $\mathcal{T}_{1\rightarrow 2} \propto \int d\phi\, e^{i(\ell_1 - \ell_2)\phi} = \delta_{\ell_1,\, \ell_2}$, through which the angular momentum, $L_z$, is conserved along the four steps which compose the cyclic schemes, and then $\ell_1 = \ell_2$. 
Beside it, by simplicity, one considers only the ground state $(\kappa = 0,\, \ell = 0) $ and the first excited state $(\kappa = 1,\, \ell = 0) $ as to engender the quantum heat engine schemes.

\section{Isomagnetic and isoenergetic cycles}

Following the framework introduced in the previous section, isomagnetic and isoenergetic quantum thermodynamic cycles can now be investigated. Modifications on the results for the efficiency coefficient, $\mathcal{N}$, under the influence of NC corrections should be expected, their effects on the quantum Carnot cycle shall be quantified and the impact of the NC deformation on the second law of thermodynamics is discussed.

\subsubsection{Isomagnetic cycle}

Let one considers the isomagnetic cycle composed by two isomagnetic and two isoentropic trajectories as depicted in Fig.~\ref{Scheme}. 
The cycle starts at the ground state (point I).
Along the trajectory $I \rightarrow II$, the work performed by the engine is null, given that the magnetic field keeps unchanged.
Consequently, all the heat exchanged with the environment is due to the energy change
from
\begin{equation}
E_{0,\, 0} (B_I) = \sigma\, \hbar\, \omega\, \sqrt{1 + \mathcal{F}_{\Phi_I}^2}, 
\end{equation}
to
\begin{equation}
E_{1,\, 0} (B_I) = 3\sigma\, \hbar\, \omega\, \sqrt{1 + \mathcal{F}_{\Phi_I}^2},
\end{equation}
with
\begin{equation}
\mathcal{F}_{\Phi_I}^2 = \frac{1}{\sigma^2 \omega^2}\left(\frac{\omega_{B_I}^2}{4} + \gamma\, \omega_{B_I} + \gamma^2\right).
\end{equation}

The heat for this process is given by,
\begin{equation}
Q_{I \rightarrow II} = \Delta E = E_{1,\, 0} (B_I) - E_{0,\, 0} (B_I) = 2\sigma\, \hbar\, \omega\, \sqrt{1 + \mathcal{F}_{\Phi_I}^2} > 0.
\label{Q12}
\end{equation}

Along the isoentropic trajectory there is no heat exchange. 
Meanwhile, the system is driven from $\ell_{B_I}$ to $\ell_{B_{III}}$, with $\ell_{_B}$ identified as the Landau Radius, $\ell_{_B} = \sqrt{\hbar/(m\,\omega_{_B})}$ \cite{Novo02}. 
It ables one to define an expansion coefficient as $\alpha = \ell_{B_{III}}/\ell_{B_I}$, such that one sets
\begin{equation}
\omega_{B_{III}} = \omega_{B_I}/\alpha^2.
\label{coef}
\end{equation}

The third peace along the trajectory corresponds again to an isomagnetic process.
The system goes from $III$ to $IV$ and one has a transition from
\begin{equation}
E_{1,\,0}(B_{III}) = 3\sigma\, \hbar\, \omega\, \sqrt{1 + \mathcal{F}_{\Phi_{III}}^2}, \end{equation}
to
\begin{equation}
E_{0,\,0}(B_{III}) = \sigma\, \hbar\, \omega\, \sqrt{1 + \mathcal{F}_{\Phi_{III}}^2},
\end{equation}
for which, by using (\ref{coef}), one obtains
\begin{equation}
\mathcal{F}_{\Phi_{III}}^2 = \frac{1}{\sigma^2 \omega^2}\left(\frac{\omega_{B_I}^2}{4\alpha^4} + \frac{\gamma\, \omega_{B_I}}{\alpha^2} + \gamma^2\right).
\end{equation}
Thus the heat exchanged with the environment is thus given by
\begin{equation}
Q_{III \rightarrow IV} = \Delta E = E_{0,\,0}(B_{III}) - E_{1,\,0}(B_{III}) = -2\sigma\, \hbar\, \omega\, \sqrt{1 + \mathcal{F}_{\Phi_{III}}^2} < 0.
\label{Q34}
\end{equation}
Finally, to close the cycle, another isoentropic is considered.

\begin{figure}
\hspace{-1.5 cm}
\centering
\includegraphics[scale=0.40]{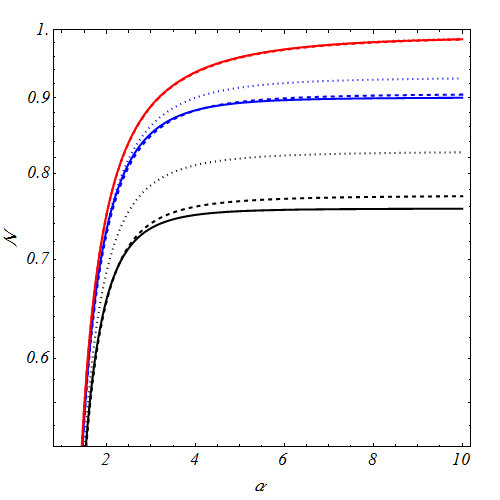}
\includegraphics[scale=0.40]{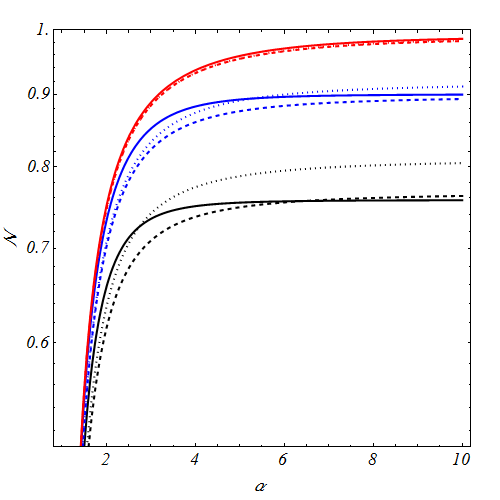}
\caption{\small (Color online) Efficiency, $\mathcal{N}$, of the isomagnetic cycle for $N_{\Phi_1}^{(0)} = 4$ (black lines), $N_{\Phi_1}^{(0)} = 10$ (blue lines) and $N_{\Phi_1}^{(0)} = 100$ (red lines) as a function of the expansion coefficient $\alpha$, for $\theta \eta = 0$ (solid lines), $\theta \eta = 0.1$ (dashed lines), $\theta \eta = 0.5$ (dotted lines) and for $\gamma = 0.1$ (left plot) and $\gamma = 0.5$ (right plot). One has considered $\hbar = 1$ and $\omega = 1$.}
\label{RendIsoMag}
\end{figure}

By computing the efficiency coefficient,
\begin{equation}
\mathcal{N} = 1 - \left|\frac{Q_{III \rightarrow IV}}{Q_{I \rightarrow II}}\right|,
\end{equation}
one can thus measure the impact of the NC QM on such quantum engine scheme.
Fig.~\ref{RendIsoMag} shows the results for $\mathcal{N}$ for a quantum heat engine along the isomagnetic cycle, in terms of the redefined parameters, $N_{\Phi_1}^{(0)} = \omega_{B_I}/2\omega$, and $\alpha$.

By focusing on the entropy generation due to the processes involved in the isomagnetic cycle, according to Ref.~\cite{Beretta}, for the Carnot and Otto-like cycles for a two-level system, the entropy balance results into the following scheme
\begin{equation}
d S = \frac{\delta Q^{\longleftarrow}}{T_H} + \delta S_{gen},
\end{equation}
where $\delta Q^{\longleftarrow}$ is the heat absorbed by the system and $T_H$ is the temperature of the hot bath.
Considering the terms on the r.h.s. of the above equation, the first one corresponds to the increasing entropy due to the heat interaction, and the second one to the increasing entropy to the internal dynamics (relaxation, decoherence). By imposing and identifying the work performed by the system just as a function of the energy gap, the expression for $\delta S_{gen}$ is computed to be given by \cite{Beretta}
\begin{equation}
\delta S_{gen} = \frac{\delta Q^{\longleftarrow}}{T} - \frac{\delta Q^{\longleftarrow}}{T_H},
\label{entropy}
\end{equation}
where $T$ is the temperature of the system, and one notices that for an isoentropic process (isothermal process) $\delta S_{gen} = 0$.

For the isomagnetic processes (or the so-called isoenergy gap processes in \cite{Beretta}), which is engendered by thermal contacts with reservoirs at temperatures $T_{high}$ and $T_{low}$, one notice from (\ref{entropy}) that they circumvent the entropy generation due to the internal dynamics, namely due to an irreversible process. However, it could be recovered, in principle, by a sequence of infinitesimal contacts with an infinite number of hot and cold baths covering the temperature of the system and the temperatures of the reservoir, $T_{high}$ and $T_{low}$, as mentioned in \cite{Beretta}. This should guarantee a theoretical formulation of our isomagnetic cycle which includes essentially two isomagnetic and two isoentropic trajectories to complete the cycle. 
Finally, from the experimental perspective, according to Ref. \cite{Scully01}, one could consider the adaptation of a laser-maser system coupled to an Otto-like engine as to effectively perform the cycle - in this case, the maser acts as the removal mechanism of incoherent energy (``heat") whereas the laser acts as the provider of coherent energy mechanism (``work") performed by the quantum engine.

\subsubsection{Isoenergetic cycle}

The isoenergetic cycle \cite{Bender2000,Bender,Novo02} is composed by two isoenergetic and two isoentropic trajectories as depicted in Fig.~\ref{Scheme}.
Unlike the isomagnetic process, it can be engendered through a careful adjust of the temperature when varying the work parameter (magnetic field).
That is, also unlike an isothermal process, the bath temperature changes along the isoenergetic trajectory in a corresponding fine-tuning with the intensity of the external magnetic field.
One assumes \cite{Bender2000} that the quantum system must be in contact with a heat bath along the isoenergetic trajectories in order to maintain the Hamiltonian expectation value of the system unchanged.
Since one varies the intensity of the external magnetic field, which drives the system from one to another energy level, the heat bath provides the condition for a quasi-static heat transference as to keep the Hamiltonian expectation value constant.
Along the isoenergetic expansion, the Landau Radius, described as a function of the magnetic field, $\ell_{_B} = \sqrt{\hbar/(m\,\omega_{_B})}$, where $\omega_{_B} \equiv \omega(B)$, increases too slowly, while the system is kept in contact with the thermal bath \cite{Bender2000, Bender}, which constrains the level transition to an isoenergetic trajectory.
From the experimental perspective, even being considered a current research topic affected by a {\em pletora} of quantum phenomena as entanglement, decoherence, etc, one suggests \cite{Beretta,Scully01} the use of a maser-laser mechanism to implement the simultaneous heat and work interaction -- through a smooth continuous change of the magnetic field, as it is performed to realize the isotherms of Carnot cycles -- sheds some light in the possibility of engendering a similar mechanism which accounts for a fine-tuning between temperature and magnetic field intensity, as to produce an isoenergetic process.

In this case, before effectively engendering the quantum cycle, it has been more convenient to compute the work and heat exchange along each independent trajectory.
Since the magnetic field intensity is the unique external driver of the system, the overall change in the averaged energy is given by \cite{Novo02}
\begin{eqnarray}
d E  &=& \sum_{\kappa,\, \ell} p_{\kappa,\, \ell}(B) d E_{\kappa,\, \ell} (B) + \sum_{\kappa,\, \ell} dp_{\kappa,\, \ell}(B) E_{\kappa,\, \ell}(B),
\label{varE}
\end{eqnarray}
where $p_{\kappa,\, \ell}(B)$ corresponds to the probability for the system to be found in a quantum state $\psi_{\kappa,\, \ell}(B)$. 
First and second terms of the right-hand side of Eq.~(\ref{varE}) reads respectively the isoentropic and the isoenergetic process.

The isoentropic trajectory exhibits a constant probability coefficient, $p_{\kappa,\, \ell}(B)$, since the infinitesimal work along the process is computed according to
\begin{equation}
dW = - M d B,
\end{equation}
where $M = -(\partial E/\partial B)$ is the magnetization considered for bringing the magnetic field intensity from $B = B_a$ to $B = B_b$, which allows for obtaining \cite{Novo02}
\begin{equation}
W_{a \rightarrow b} = \sum_{\kappa,\, \ell} p_{\kappa,\, \ell}(B_a)\, [E_{\kappa,\, \ell}(B_b) - E_{\kappa,\, \ell}(B_a)].
\end{equation}

The isoenergetic trajectory exhibits a constant energy coefficient, $E_{\kappa,\, \ell}$.
According to the first law of thermodynamics, along the isoenergetic process, one has
\begin{equation}
\Delta E = W_{a \rightarrow b} + Q_{a \rightarrow b} = 0,
\end{equation}
which implies into $Q_{a \rightarrow b} = - W_{a \rightarrow b} $.
Thus, one can compute the heat exchanged between the system and the environment in terms of
\cite{Novo02}
\begin{equation}
Q_{a \rightarrow b} = - \sum_{\kappa,\, \ell} \int_{B_a}^{B_b} E_{\kappa,\, \ell} \frac{d p_{\kappa,\, \ell} (B)}{d B} dB,
\label{label01}
\end{equation}
where $B$ changes from $B_a$ to $B_b$ and $p_{\kappa,\, \ell}(B)$ satisfies the normalization condition
\begin{equation}
\mbox{Tr} \hat{\rho} = \sum_{\kappa,\, \ell} p_{\kappa,\, \ell}(B) = 1,
\label{norm}
\end{equation}
where $\hat{\rho}$ is the density matrix operator.

Since, for the isoenergetic process, the energy satisfies the condition expressed by
\begin{equation}
\sum_{\kappa,\, \ell} p_{\kappa,\, \ell}(B) E_{\kappa,\, \ell}(B) = \sum_{\kappa,\, \ell}p_{\kappa,\, \ell} (B_a) E_{\kappa,\, \ell} (B_a),
\end{equation}
and following the isoenergetic trajectories, with the condition from (\ref{norm}), one has
\begin{equation}
p_{0,\, 0}(B) = \frac{E_{1,\, 0}(B_a) - E_{1,\, 0}(B)}{E_{0,\, 0}(B) - E_{1,\, 0}(B)} + \frac{E_{0,\, 0}(B_a) - E_{1,\, 0}(B_a)}{E_{0,\, 0}(B) - E_{1,\, 0}(B)}\,p_{0,\, 0}(B_a),
\end{equation}
the heat exchanged along the isoenergetic trajectory can be analytically given by
\begin{equation}
Q_{a \rightarrow b} = \lbrace E_{1,\, 0}(B_a) + [E_{0,\, 0}(B_a) - E_{1,\, 0}(B_a)] p_{0,\, 0}(B_a)\rbrace\times \mbox{ln}\left[\frac{E_{0,\, 0}(B_b) - E_{1,\, 0}(B_b)}{E_{0,\, 0}(B_a) - E_{1,\, 0}(B_a)}\right].
\label{heat01}
\end{equation}

Turning back to the closed cycle, it starts with an isoenergetic expansion from $\ell_{B_1}$ to $\ell_{B_2}$, with
\begin{equation}
p_{0,\, 0}(B_1) = 1, \quad \mbox{and} \quad \ell_{B_2} = \alpha_1 \ell_{B_1},
\end{equation}
where $\alpha_1$ is an expansion coefficient and
\begin{equation}
E_{0,\,0}(B_1) = \sigma\, \hbar\, \omega \sqrt{1 + \mathcal{F}_{\Phi_1}^2}, 
\end{equation}
\begin{equation}
E_{1,\,0}(B_2) = 3\sigma\, \hbar\, \omega \sqrt{1 + \mathcal{F}_{\Phi_2}^2},
\end{equation}
where
\begin{eqnarray}
\mathcal{F}_{\Phi_1}^2 &= &\frac{1}{\sigma^2 \omega^2}\left(\frac{\omega_{B_1}^2}{4} + \gamma\, \omega_{B_1} + \gamma^2\right),\\
\mathcal{F}_{\Phi_2}^2 &=& \frac{1}{\sigma^2 \omega^2}\left(\frac{\omega_{B_1}^2}{4\alpha_1^4} + \frac{\gamma\, \omega_{B_1}}{\alpha_1^2} + \gamma^2\right).
\end{eqnarray}
The isoenergetic condition sets $E_{0,\,0}(B_1)$ equals to $E_{1,\,0}(B_2)$, and thus the expansion coefficient $\alpha_1$, results into
\begin{equation}
\alpha_1 = \left[\frac{3 \omega^2 N_{\Phi_1}^{(0)^2}}{\sqrt{\omega^2N_{\Phi_1}^{(0)^2}(\gamma^2 + 2\omega \gamma N_{\Phi_1}^{(0)} + \omega^2(N_{\Phi_1}^{(0)^2} - 8\sigma^2))} - 3\omega \gamma N_{\Phi_1}^{(0)}}\right]^{1/2}.
\end{equation}

For the following isoentropic expansion, one has
\begin{equation}
p_{1,\, 0}(B_2) = 1, \quad \mbox{and} \quad \ell_{B_3} = \alpha \ell_{B_2} = \alpha\, \alpha_1\, \ell_{B_1},
\end{equation}
and no heat exchange.

The system then undergoes an isoenergetic compression such that
\begin{equation}
\ell_{B_3} \rightarrow \ell_{B_4} < \ell_{B_3}, \quad \mbox{with} \quad \ell_{B_4} = \alpha_3\, \ell_{B_3} = \alpha_1\, \alpha_3\, \alpha\, \ell_{B_1}.
\end{equation}

Analogous to what happens along the isoenergetic expansion, one also has
$E_{1,\,0}(B_3)$ equals to $E_{0,\,0}(B_4)$, which implies into
\begin{equation}
\sqrt{1 + \mathcal{F}_{\Phi_4}^2} = 3\sqrt{1 + \mathcal{F}_{\Phi_3}^2},
\label{eq3}
\end{equation}
where
\begin{eqnarray}
\mathcal{F}_{\Phi_3}^2 &=& \frac{1}{\sigma^2 \omega^2}\left(\frac{\omega_{B_1}^2}{4(\alpha_1\,\alpha)^4} + \frac{\gamma\, \omega_{B_1}}{(\alpha_1\,\alpha)^2} + \gamma^2\right),\\
\mathcal{F}_{\Phi_4}^2 &=& \frac{1}{\sigma^2 \omega^2}\left(\frac{\omega_{B_1}^2}{4(\alpha_1\,\alpha\,\alpha_3)^4} + \frac{\gamma\, \omega_{B_1}}{(\alpha_1\,\alpha\,\alpha_3)^2} + \gamma^2\right).
\end{eqnarray}

By solving Eq.~(\ref{eq3}) for $\alpha_3$, one has
\begin{equation}
\alpha_3 = \left[\frac{\omega^2 N_{\Phi_1}^{(0)^2}}{\sqrt{\omega^2N_{\Phi_1}^{(0)^2}(18\omega
_d(\alpha \alpha_1)^2 \gamma N_{\Phi_1}^{(0)} + 9\omega^2 N_{\Phi_1}^{(0)^2} + (\alpha \alpha_1)^4(9\gamma^2 + 8\sigma^2\omega^2))} - \omega (\alpha \alpha_1)^2 \gamma N_{\Phi_1}^{(0)}}\right]^{1/2}.
\end{equation}

Finally, the Eq. (\ref{heat01}) for the heat exchanged with the environment provides the efficiency of the quantum heat engine working along the isoenergetic cycle, as
\begin{equation}
\mathcal{N}(N_{\Phi_1}^{(0)}, \sigma, \gamma, \alpha) = 1 - 3\frac{\Theta(\alpha \alpha_1)}{\Theta(1)}\frac{\mbox{ln}\left[\frac{\Theta(\alpha \alpha_1 \alpha_3)}{\Theta(\alpha \alpha_1)}\right]}{\mbox{ln}\left[\frac{\Theta(1)}{\Theta(\alpha_1)}\right]},
\label{RendIE}
\end{equation}
where one has identified
\begin{equation}
\Theta (x) = \sqrt{1 + \frac{1}{\sigma^2 \omega^2}\left(\frac{\omega_{B_1}^2}{4 x^4} + \frac{\gamma\, \omega_{B_1}}{x^2} + \gamma^2\right)}.
\label{Theta}
\end{equation}

\begin{figure}
\hspace{-1.5 cm}
\centering
\includegraphics[scale=0.40]{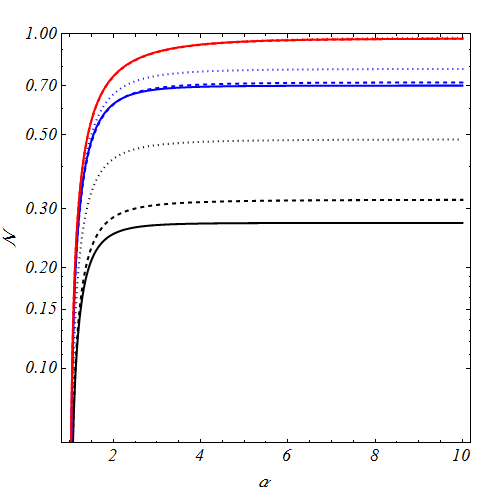}
\includegraphics[scale=0.40]{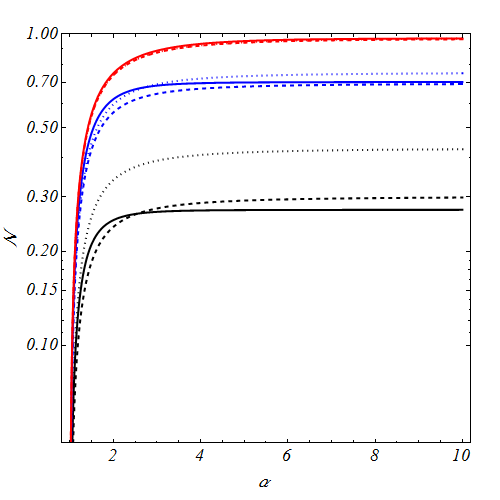}
\caption{\small (Color online) Efficiency, $\mathcal{N}$, of the isoenergetic cycle for $N_{\Phi_1}^{(0)} = 4$ (black lines), $N_{\Phi_1}^{(0)} = 10$ (blue lines) and $N_{\Phi_1}^{(0)} = 100$ (red lines) as a function of the expansion coefficient $\alpha$, for $\theta \eta = 0$ (solid lines), $\theta \eta = 0.1$ (dashed lines), $\theta \eta = 0.5$ (dotted lines) and for $\gamma = 0.1$ (left plot) and $\gamma = 0.5$ (right plot). One has considered $\hbar = 1$ and $\omega = 1$.}
\label{RendIsoEnerg}
\end{figure}

Fig.~\ref{RendIsoEnerg} shows the efficiency for the isoenergetic cycle driven by the NC parameter for different values of $N_{\Phi_1}^{(0)}$.
Considering the asymptotic limit for very large magnetic fields, namely when $N_{\Phi_1}^{(0)}$ goes to $\infty$, one notices a typical behavior given by
\begin{equation}
\mathcal{N}(N_{\Phi_1}^{(0)}, \sigma, \gamma, \alpha) \rightarrow 1- 1/\alpha^2,
\label{limit}
\end{equation}
which coincides with the results from Refs. \cite{Novo02,Novo03}.
It means that the NC parameters do not affect the maximal efficiency of the quantum heat engine working under an isoenergetic cycle. 

The limit of $\mathcal{N}$ for the vanishing value of $N_{\Phi_1}^{(0)}$ also does not exhibit any residual effect from NC QM. 
Although $\gamma$ works as an effective magnetic field, it does not work as an extensive thermodynamic quantity since, 
from  Eq.~(\ref{Theta}), $\gamma^2$ indeed does not depend on $\alpha$.

From Figs.~ \ref{RendIsoMag} and \ref{RendIsoEnerg}, respectively for isomagnetic and isoenergetic cycles, one notices that varying $\gamma$ from $0.1$ to $0.5$ suppress the NC imprints on the efficiency coefficient, $\mathcal{N}$. 
Both $\gamma$ and $\omega_{_B}$ indeed amplify the magnetic field effect on the quantum heat engine. Otherwise, as reported by Eq.~(\ref{limit}), as the intensity of the magnetic fields increases, the contribution due to NC corrections are highly suppressed. For instance, in case of having $N_{\Phi_1}^{(0)} = 100$, the system shall reach the highest efficiency value just for $\theta\eta = 0$.

\subsubsection{The quantum Carnot cycle and NC effects}

NC signatures on quantum heat engines could be expected to be relevant for the quantum Carnot cycle, which can be investigated in the same context as it was performed in Refs. \cite{Novo02}.
An engine working under the quantum Carnot cycle through the variation of the intensity of the external magnetic field  has the efficiency unchanged by the magnetic field and can be simply computed in terms of the environment temperatures \cite{Novo02} as to give
\begin{equation}
\mathcal{N}^C = 1 - \frac{T_C}{T_H},
\label{EficCarnot}
\end{equation}
where $T_C$ and $T_H$ are the temperature of the cold and hot reservoir respectively.

This result indicates a universal maximal efficiency for any engine operating under the Carnot cycle, even in the NC QM regime. By including the NC effects as to describe an effective magnetic field through
$\mathcal{B} = \mathcal{B}(\theta, \eta)$ changes the scale of the original magnetic field, in spite of not modifying the the efficiency of the Carnot cycle. This reinforces the validity of the second law of thermodynamics which imposes universal limit for transforming a certain quantity of heat into profitable work.

\section{Conclusions}

Two theoretically relevant examples of quantum heat engines involving isomagnetic and isoenergetic cycles have been considered in a scenario of NC QM. 
For both cases, analytical expressions for the efficiency of quantum heat engines as function of the NC parameters and typical driving parameters have been obtained.
Generically, our results point to that the support of a deformed Heisenberg algebra produces an effective external magnetic field which increases the efficiency of the quantum engines here investigated.

For the isoenergetic cycle, one has found that the magnetic field asymptotic limit leads to a universal result for the efficiency coefficient, $\mathcal{N} \sim 1 - 1/\alpha^2$, which does not depend on the NC effects, and which is in agreement with previously reported results \cite{Novo02,Bender2000, Bender,Novo03}. 
More generically speaking, for the isoenergetic cycle, the suppression of the intensity of the magnetic field is also followed by a suppression of the NC effects.
%The validity of this limit even considering a deformed Heisenberg algebra suggests it may represent a universal upper limit for the isoenergetic cycle. Moreover, when taking the limit of vanishing the external magnetic field one can not observe any NC signature. This occurs because the NC effect is a fixed parameter and not a new degree of freedom, just acting as recalibration in the original magnetic field.

Still concerning the inclusion of NC corrections, one of the most relevant results is that, for the quantum Carnot cycle, even considering the NC extension of the QM, the efficiency coefficient, $\mathcal{N}^C$, does not depend on the NC parameters, and reproduce the results for ordinary QM.
This remarkable universal character reinforces the conceptual robustness of the second law of thermodynamics. Finally, once the NC corrections have the behavior of an effective magnetic field $\mathcal{B} = \mathcal{B}(\theta, \eta)$, this indicates that such an effect may be simulated in some quantum heat engines operating under the isomagnetic or isoenergetic schemes.

\section*{Appendix I: The SW map and the effective magnetic external field}

The NC theory acts in a $2D$-dimensional phase-space where time is identified as a commutative variable, and coordinate and momentum components obey a NC algebra.
In general, NC generalizations of QM are more conveniently formulated through the Weyl-Wigner-Groenewold-Moyal (WWGM) formalism for QM \cite{Groenewold,Moyal,Wigner}.
However, for the most relevant Hamiltonian systems, the phase-space NC extensions of QM can be effectively implemented through the deformed Heisenberg-Weyl algebra from Eq.~(\ref{Ape001}), where $\eta_{ij}$ and $\theta_{ij}$ are invertible antisymmetric real constant ($D \times D$) matrices, and one can define the matrix
\begin{equation}
\Sigma_{ij} \equiv \delta_{ij} + {1\over\hbar^2}  \theta_{ik} \eta_{kj},
\label{Ape002}
\end{equation}
which is also invertible if $\theta_{ik}\eta_{kj} \neq -\hbar^2 \delta_{ij}$. 

The connection between NC and ordinary QM in the context of a quantum field theory allows one to implement the Seiberg-Witten (SW) map such that the NC parameters could be read in terms of effective magnetic external field components in an ordinary QM scenario \cite{Bernardini01,Rosenbaum}. 

The SW map corresponds to a linear transformation which maps the algebra from Eq.~(\ref{Ape001}) into the standard commutation relations of the QM.
It ensures to the NC algebra an operational representation in terms of the Hilbert space of ordinary QM. Thus, the NC parameters in Hamiltonian (polynomial) systems work effectively as components of an additional external magnetic field, as pointed out along this work.

\section*{Appendix II: Efficiency for the opposite orientation of the magnetic field}

\begin{figure}
%\hspace{-1.5 cm}
\centering
\includegraphics[scale=0.40]{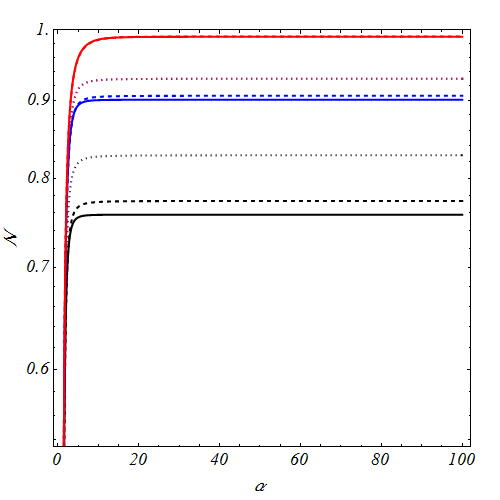}
\includegraphics[scale=0.40]{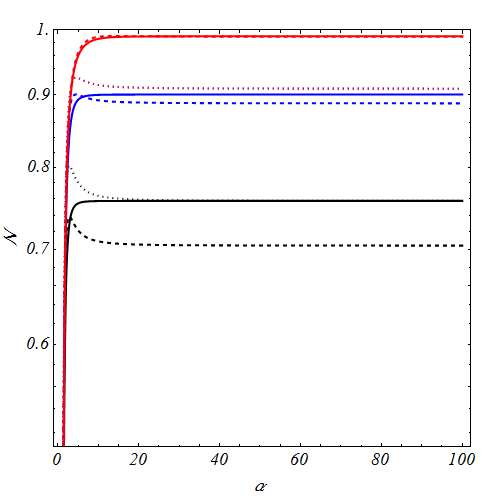}

\includegraphics[scale=0.40]{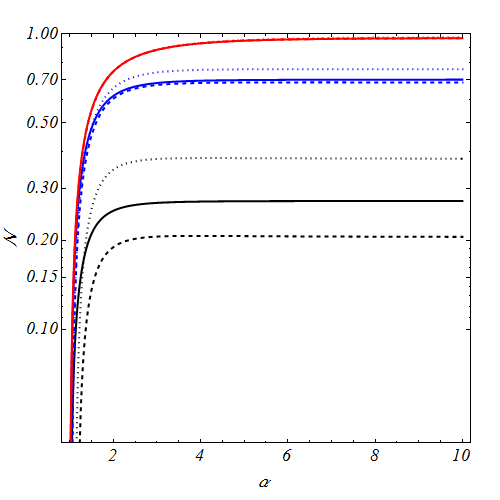}
\includegraphics[scale=0.40]{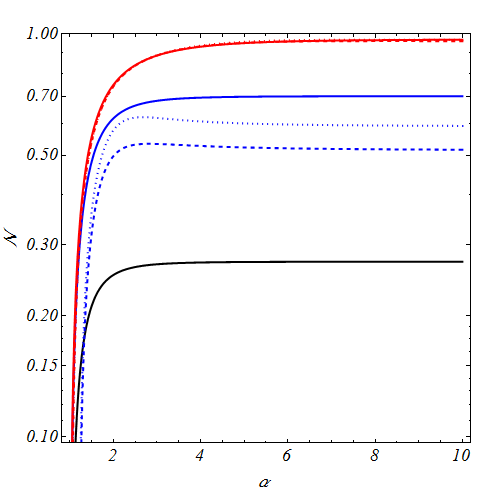}
\caption{\small (Color online) Efficiency, $\mathcal{N}$, of the isomagnetic (first row) and of the isoenergetic (second row) cycles for $N_{\Phi_1}^{(0)} = 4$ (black lines), $N_{\Phi_1}^{(0)} = 10$ (blue lines) and $N_{\Phi_1}^{(0)} = 100$ (red lines) as a function of the expansion coefficient $\alpha$, for $\theta \eta = 0$ (solid lines), $\theta \eta = 0.1$ (dashed lines), $\theta \eta = 0.5$ (dotted lines) and for $\gamma = 0.1$ (left plot) and $\gamma = 0.5$ (right plot) and considering an opposite orientation of the magnetic field, such that $B \rightarrow -B$. One has considered $\hbar = 1$ and $\omega = 1$.}
\label{Anexo}
\end{figure}

Here, one illustrates the efficiency for the isomagnetic and isoenergetic cycles but with the opposite orientation of the magnetic field, in order to clarify that the NC parameters, $\gamma$, can also decrease the efficiency of the cycles, as depicted in Figure \ref{Anexo}. Notice that the limit for high magnetic field intensity remains the same.

It can be argued about a value of magnetic field intensity, or the frequency associated with $B$, $\omega_B$, which is necessary to vanish the effect of the NC parameter, $\gamma$. By considering $\omega_B \rightarrow -\omega_B$ in Eq. (\ref{Eingelvalues}), one obtains
\begin{equation}
E_{\kappa,\, \ell} = \sigma\, \hbar\, \omega\,\sqrt{1 + \mathcal{F}_\Phi^2}\,(2\kappa + |\ell| + 1) - \hbar \left(-\frac{\omega_{_B}}{2} + \gamma\right)\ell,
\label{Eingelvalues02}
\end{equation}
where $\mathcal{F}_\Phi^2$ is now given by
\begin{equation}
\mathcal{F}_\Phi^2 = \frac{1}{\sigma^2 \omega^2}\left(\frac{\omega_{_B}^2}{4} - \gamma\omega_{_B} + \gamma^2\right),
\end{equation}

By imposing that $\mathcal{F}_\Phi^2 = 0$ one obtains the condition that $\omega_B = 2\gamma$ which is the value of the associated frequency of the magnetic field such that it cancels the NC effects.

{\em Acknowledgments - The work of AEB is supported by the Brazilian Agencies FAPESP (grant 15/05903-4) and CNPq (grant 300809/2013-1 and grant 440446/2014-7).}

\newpage

\begin{thebibliography}{99}
\bibitem{Novo01}
J. Gemmer, M. Michel, and G. Mahler, {\em Quantum Thermodynamics} (Springer, Berlin-Heidelberg, 2009).
\bibitem{Novo02}
E. Mu\~noz and F. J. Peña, Phys. Rev. {\bf E89}, 052107 (2014).
%\bibitem{Novo03B}
%E. Mu\~noz, Z. Barticevic and M. Pacheco, Phys. Rev. {\bf B71}, 165301 (2005).
\bibitem{Novo03}
H. T. Quan, Phys. Rev. {\bf E79}, 041129 (2009).
\bibitem{Novo04}
C. Bustamante, J. Liphardt, and F. Ritort, Phys. Today {\bf 54}, (7) 43 (2005).
\bibitem{3000}
O. Abah, J. Roßnagel, G. Jacob, S. Deffner, F. Schmidt- Kaler, K. Singer and E. Lutz, Phys. Rev. Lett. {\bf 109}, 203006 (2012). 
\bibitem{Novo00}
B. Gardas and S. Deffner, Phys. Rev. {\bf E92}, 042126 (2015).
\bibitem{5000}
 D. Gelbwaser-Klimovsky and A. Aspuru-Guzik, J. Phys. Chem. Lett. {\bf 6}, 3477 (2015).
\bibitem{6000}
E. J. O'Reilly and A. Olaya-Castro, Nat. Commun. 5, 3012 (2014).
\bibitem{7000}
E. Romero {\em et al.},  Nature Phys. {\bf 10}, 676 (2014).
\bibitem{1300}
H. Li, J. Zou, W.-L. Yu, B.-M. Xu, and B. Shao, Eur. Phys. J. {\bf D86}, 67 (2013).
\bibitem{1400}
J. Roßnagel, O. Abah, F. Schmidt-Kaler, K. Singer and E. Lutz, Phys. Rev. Lett. {\bf 112}, 030602 (2014).
\bibitem{1000}
M. O. Scully, Phys. Rev. Lett. {\bf 88}, 050602 (2002).
\bibitem{1100}
M. O. Scully, M. S. Zubairy, G. S. Agarwal and H. Walther, Science {\bf 299}, 862 (2003).
\bibitem{1200}
M. O. Scully, K. R. Chapin, K. E. Dorfman, M. B. Kim and A. Svidzinsky, Proc. Natl. Acad. Sci. USA {\bf 108}, 15097 (2011).
\bibitem{Novo0B}
R. Kosloff, Entropy {\bf 15}, 2100 (2013).
\bibitem{Bernardini01}
A. E. Bernardini and O. Bertolami, Phys. Rev. {\bf A88}, 012101 (2013).
\bibitem{Catarina}
C. Bastos, O. Bertolami, N. C. Dias and J. N. Prata, J. Math. Phys. {\bf 49}, 072101 (2008). 
 \bibitem{Catarina001}
C. Bastos, O. Bertolami, N. C. Dias and J. N. Prata, Phys. Rev. {\bf D78}, 023516 (2008).
\bibitem{Catarina002}
C. Bastos, O. Bertolami, N. C. Dias and J. N. Prata, Int. J. Mod. Phys. {\bf A24}, 2741 (2009). 
\bibitem{Bernardini13B}
C. Bastos, A. E. Bernardini, O. Bertolami, N. C. Dias and J. N. Prata, Phys. Rev. {\bf D88}, 085013 (2013).
\bibitem{Bernardini13B2}
C. Bastos, A. E. Bernardini, O. Bertolami, N. C. Dias and J. N. Prata, Phys. Rev. {\bf D93}, 104055 (2016).
\bibitem{2015} 
J. F. G. Santos, A. E. Bernardini and C. Bastos, Physica {\bf A438}, 340 (2015).
\bibitem{2016} 
J. F. G. Santos and A. E. Bernardini, Physica {\bf A 445}, 75 (2016).
\bibitem{Bernardini13C}
C. Bastos, A. E. Bernardini, O. Bertolami, N. C. Dias and J. N. Prata, Phys. Rev.{\bf A89}, 042112 (2014). 
\bibitem{Bernardini13D}
C. Bastos, A. E. Bernardini, O. Bertolami, N. C. Dias and J. N. Prata, Phys. Rev. {\bf D90}, 045023 (2014).
\bibitem{Bernardini13E}
C. Bastos, A. E. Bernardini, O. Bertolami, N. C. Dias and J. N. Prata, Phys. Rev. {\bf D91}, 065036 (2015).
\bibitem{Gamboa} 
J. Gamboa, M. Loewe, F. Mendez and J. C. Rojas, Mod. Phys. Lett. {\bf A16}, 2075 (2001).
\bibitem{Rosenbaum}
M. Rosenbaum, J. D. Vergara and L. R. Juarez, Phys. Lett. {\bf A367}, 1 (2007).
\bibitem{Nekrasov01}
M. R. Douglas and N. A. Nekrasov, Rev. Mod. Phys. {\bf 73}, 977 (2001).
\bibitem{Prange}
R. Prange and S. Girvin {\em The Quantum Hall Effect} (Springer, New York, 1987).
\bibitem{Bertolami01}
O. Bertolami, J. G. Rosa, C. M. L. de Arag\~ao, P. Castorina and D. Zappal\`a, Phys. Rev. {\bf D72}, 025010 (2005).
\bibitem{Bertolami02}
O. Bertolami, J. G. Rosa, C. Arag\~ao, P. Castorina and D. Zappal\`a, Mod. Phys. Lett. {\bf A 21}, 795 (2006).
\bibitem{Banerjee}
R. Banerjee, B. D. Roy and S. Samanta, Phys. Rev. {\bf D74}, 045015 (2006).
\bibitem{Connes}
A. Connes, M. R. Douglas and A. Schwarz, JHEP {\bf 02}, 003 (1998).
\bibitem{Douglas}
M. R. Douglas and C. Hull, JHEP {\bf 02}, 008 (1998);
V. Schomerus, JHEP {\bf 9906}, 030 (1999).
\bibitem{Seiberg}
N. Seiberg and E. Witten, JHEP {\bf 9909}, 032 (1999).
\bibitem{Jacak}
L. Jacak, P. Hawrylak, and A. Wójs, Quantum Dots (Springer-Verlag, 1998).
\bibitem{Pal}
B.K. Pal, B. Roy, B. Basu, Phys. Lett. {\bf A42}, 374 (2010).
\bibitem{Beretta}
G. P. Beretta, Exp. Phys. Lett, {\bf 99}, 20005 (2012).
\bibitem{Scully01}
M. O. Scully, Phys. Rev. Lett. 050602 (2002).
\bibitem{Bender2000}
C. M. Bender, D. C. Brody and B. K. Meister,  J. Phys. A: Math. Gen. {\bf 33}, 4427 (2000).
\bibitem{Bender}
C. M. Bender, D. C. Brody and B. K. Meister, Proc. R. Soc. {\bf A458}, 1519 (2002).
\bibitem{Groenewold} 
H. Groenewold, Physica {\bf 12}, 405 (1946).
\bibitem{Moyal}  
J. Moyal, Proc. Camb. Phil. Soc. {\bf 45}, 99 (1949).
\bibitem{Wigner}
E. Wigner, Phys. Rev. {\bf 40}, 749  (1932).
\end{thebibliography}
\end{document}